\documentclass[12pt,twoside]{article}\usepackage[margin=1in,centering]{geometry}
 \newcommand\ttl {Notes for Miscellaneous Lectures}
 \newcommand\aut {Leonid A.~Levin}
 \usepackage[unicode]{hyperref} \usepackage[utf8]{inputenc}
 \usepackage{microtype,amsmath,amssymb,amsthm,bm,mmap,xcolor} 
\begin{document}

 \frenchspacing \newcommand\hreff[1] {{\footnotesize \url {https://#1}}}
 \renewcommand\smile{\mbox{:-)}} \newtheorem{lemma} {Lemma}
 \newcommand\emm[1]{{\ensuremath{#1}}} \newcommand\trm[1]{{\bf\em #1}}
 \newcommand\edf{{\raisebox{-2pt}{$\,\stackrel{\mbox{\tiny df}}=\,$}}}
 \newcommand\ceil[1]{{\lceil#1\rceil}} \newcommand\drp[1]{}

 \newcommand\ex\exists \newcommand\all\forall
 \newcommand\then\Rightarrow \renewcommand\iff\Leftrightarrow

 \renewcommand\a{{\emm\alpha}} \renewcommand\b{{\emm\beta}}
 \newcommand\e\varepsilon \renewcommand\l{\lambda} \newcommand\w\omega
 \newcommand\W{{\mathbf\Omega}} \newcommand\g{{\emm\gamma}}
 
 \newcommand\N{{\mathbb N}}\newcommand\R{{\mathbb R}}\newcommand\Z{{\mathbb Z}}
 \renewcommand\L{{\mbox{\bf L}}} \newcommand\Kt{{\mathbf Kt}}
 \newcommand\M{{\mathbf M}} \newcommand\T{{\mathbf T}}
 \newcommand\m{{\mathbf m}} \newcommand\K{{\mathbf K}}
 \renewcommand\d{{\mathbf d}} \newcommand\St{{\mathbf S}}

 \author{\aut\\ (\hreff {www.cs.bu.edu/fac/lnd/}) \\ Boston University\thanks
 {College of Arts and Sciences, Computer Science department, Boston,
 MA 02215, USA}} \title{\ttl}\date{}\maketitle \begin{abstract}

   Here I share a few notes I used in various
  course lectures, talks, etc. Some may be just calculations that in
  the textbooks are more complicated, scattered, or less specific;
  others may be simple observations I found useful or curious.

 \end{abstract} \tableofcontents
 \vfill\noindent Copyright \textcircled{c} $\number\year$ by the author.
 \hfill Last revised: \today.

 \newpage \section [Nemirovski Estimate of Common Mean of Distributions]
 {Nemirovski Estimate of Common Mean of\\
              Arbitrary Distributions with Bounded Variance}

 The popular Chernoff bounds\footnote
 {First studied by S.N. Bernstein: {\em Theory of Probability.}, Moscow,
1927. Tightened by Wassily Hoeffding in: Probability inequalities for
sums of bounded random variables, J.Am.Stat.Assoc. 58(301):13-30, 1963.}
 assume severe restrictions on distribution: it must be cut-off,
 or vanish exponentially, etc. In [Nemirovsky Yudin]\footnote
 {A.S.Nemirovsky, D.B.Yudin. {\em Problem Complexity
 and Method Efficiency in Optimization.} Wiley, 1983.}
 an equally simple bound uses no conditions at all
 beyond independence and known bound on variance.
 It is not widely used because it is not explained anywhere
 with an explicit tight computation. I offer this version:

\newcommand\var{{\mbox{\bf var}}}\newcommand\iv{{\mbox{\bf iv}}}
 Assume independent variables $X_i(\w)$ with the same unknown mean $m$
and known lower bounds $B_i^2$ on inverses $\iv(X_i)\edf1/\var(X_i)$ of
their variance. We estimate $m$ as $M(\w)$ with probability
$ p^\pm\edf P(\pm(M{-}m){\ge}\e)< 2^{-k}$ for $k$ close to
$\sum_i(B_i\e)^2/12$. We scale $X_i$ to set $\e{=}1$.

 {\bf Additivity.} Variance of sum of pairwise independent variables is
additive.\\ So, it grows linearly, not quadratically, with the number of
variables.\\ Weighted mean $X{=}\sum_i w_iX_i/\sum_iw_i$ shrinks the
variance. The maximal shrink is\\ with weights $w_i=\iv(X_i)$.
In this case it is {\iv} that is additive: $\iv(X)=\sum_i\iv(X_i)$.

First, we spread $X_i$ into $n$ groups, and take in each group $j$ its
$\iv(X_i)$-weighted mean $x_j(\w)$. Using additivity of {\iv} we grow groups
to get $b^2_j\edf\iv(x_j)\ge6$, to increase the sum $k$ of {\em heights}
$h_j\edf\log_2((b_j{+}b_j^{-1})/2)$. ($b^2=6$ scales precision/$\sigma=\e b$
to $\approx2.45$, makes $h>1/2$, nearly maximizing $h/b^2$. These
values can be taken below instead of $b_j^2,h_j$, for simplicity.)

For $s\subset[1,n]$, let $b_s\edf\prod_{j\in s}b_j$.
Let $L$ be the set of {\em light} $s$: with $b_s^2< b_{[1,n]}$.\\ Let $L_t$
consist of $s$ whose largest superset $s'$ in $L$ has $\|s\|{+}t$ elements.
 As $s\in L_t$ make an anti-chain (do not include each other), by
Sperner theorem, $\|L_t\|\le\binom n{\ceil{n/2}}<2^{n+1}/\sqrt{\pi(2n{+}1)}$.
 
Our $M$ is the $(\log b_j)$-weighted median of $x_j$.
 Then $\pm(M(\w){-}m)\ge1$ means\\
 $S^\pm(\w)\edf\{j:\pm (x_j{-}m)<1\}\in L$. By Cantelli's
 inequality, $p_j^\pm\edf P(j\not\in S^\pm)\le 1/(b_j^2{+}1)$.\\
 As $b_j$ are just bounds we can assume $p_j^\pm=1/ (b_j^2{+}1)$.
 If $s\in L_t$, $S^\pm(\w)=s$ has probability
     \[p_s^\pm=b_s^2/\prod_{j\le n} (b_j^2{+}1) <
 \frac{b_{s'}^2}{6^t}\frac{b_{[1,n]}}{b_{s'}^2}/\prod_{j\le n}(b_j^2{+}1)
  < 6^{-t}\prod_{j\le n} b_j/({b_j^2{+}1})= 6^{-t}2^{-(k+n)}.\]
 \[\mbox{So, } p^\pm\le\sum_{t\ge0}\sum_{s\in L_t}p_s^\pm\le (\sum_{t\ge0}
   6^{-t}) 2^{-(k+n)} 2^{n+1}/\sqrt{\pi(2n{+}1)} <2^{-k}/\sqrt{n}.\;\qed\]

 \drp{ https://math.stackexchange.com/questions/58560
       Elementary-central-binomial-coefficient-estimates
  $2^{n+1}/\binom n{\ceil{n/2}}=\sqrt{\pi(2n+1+\e/(2n+3))}, \e\in[0,1]$.
 ..bin/math: Table[{b,(Log[2,(b+1)/(2*b^.5)]/b)},{b,5.99,6.02,.0001}] }  

\newpage\section {Leftover Hash Lemma}

The following Lemma is often useful to convert a stream of symbols with
absolutely unknown (except for a lower bound on its entropy) distribution
into a source of perfectly uniform independent random bits $b\in
Z_2=\{0,1\}$.

The version I give is close to that in [HILL]\footnote
 {Johan Hastad, Russell Impagliazzo, Leonid A. Levin, Michael
Luby.\\ A Pseudorandom Generator from any One-way Function. Section 4.5.
{\em SICOMP} {\bf 28}(4):1364-1396, 1999.},
 though some aspects
are closer to that from [GL]\footnote
 {Oded Goldreich, Leonid A. Levin. A Hard-core Predicate for any One-way
Function. Sec.5. {\em STOC} 1989.}.
 Unlike [GL], I do not restrict hash functions to be linear and do
not guarantee polynomial reductions, i.e.\ I forfeit the case when the
unpredictability of the source has computational, rather than truly
random, nature. However, like [GL], I restrict hash functions only in
probability of collisions, not requiring pairwise uniform distribution.

 Let $G$ be a probability distribution on $Z_2^n$ with Renyi entropy
$-\log\sum_x G^2(x)$ $\ge m$.\\ Let $f_h(x){\in}Z_2^k$, $h{\in}Z_2^t$,
 $x{\in}Z_2^n$ be a hash function family in the sense that for each $x$,
 $y{\ne}x$\\ the fraction of $h$ with $f_h(x){=}f_h(y)$ is $\le 2^{-k}+2^{-m}$.
 Let $U^t$ be the uniform probability distribution on $Z_2^t$ and
$s=m-k-1$. Consider a distribution $P(h,a)=2^{-t} G(f^{-1}_h(a))$
 generated by identity and $f$ from $U^t\otimes G$.
 Let $\L_1(P,Q)= \sum_z|P(z)-Q(z)|$ be the $\L_1$ distance between
distributions $P$ and $Q=U^i$, $i=t+k$. It never exceeds their $\L_2$ distance
 \[\L_2(P,Q)=\sqrt{2^i\sum_z(P(z)-Q(z))^2}\;.\]

\begin{lemma}[Leftover Hash Lemma] ~
 $\L_1(P,U^i)\le\L_2(P,U^i)<2^{-s/2}\;.$ \end{lemma}

Note that $h$ must be uniformly distributed but can be reused for many
different $x$.\\ These $x$ need to be independent only of $h$, not of
each other as long as they\\ have $\ge m$ entropy in the distribution
conditional on all their predecessors.

\paragraph {Proof.} \begin {eqnarray*} (\L_2(P,U))^2 &=& 2^i\sum_{h,a} P(h,a)^2
 + 2^i\sum_z (2^{-2i}-2P(z)2^{-i}) = 2^i\sum_{h,a} P(h,a)^2 -1\\
 &=& -1+ 2^i\sum_{x,y}G(x)G(y)2^{-2t}\sum_a\|\{h\!:f_h(x)=f_h(y)=a\}\|\\
 &=& -1+ 2^{k-t}\sum_{x,y}G(x)G(y) \|\{h\!:f_h(x){=}f_h(y)\}\|\\
 &=& -1+ 2^{k-t}\left(\sum_xG(x)^2 2^t +
       \sum_{x,y\ne x}G(x)G(y) \|\{h\!:f_h(x){=}f_h(y)\}\|\right)\\
 &\le& -1+ 2^k2^{-m}+2^{k-t}(1-2^{-m})2^t(2^{-k}+2^{-m}) < 2^{-s}\;.
 \end{eqnarray*}\qed

\newpage\section {Disputed Ballots and Poll Instabilities}

Here is another curious example of advantages of quadratic norms.

The ever-vigilant struggle of major parties for the heart of the median
voter makes many elections quite tight. Add the Electoral College system
of the US Presidential elections and the history may hang on a small
number of ballots in one state. The problem is not in the randomness of
the outcome. In fact, chance brings a sort of fair power sharing
unplagued with indecision: either party wins sometimes, but the country
always has only one leader. If a close race must be settled by dice, so
be it. But the dice must be trusty and immune to manipulation!

Alas, this is not what our systems assure. Of course, old democratic
traditions help avoiding outrages endangering younger democracies, such
as Ukraine. Yet, we do not want parties to compete on tricks that may
decide the elections: appointing partisan election officials or judges,
easing voter access in sympathetic districts, etc. Better to make
the randomness of the outcome explicit, giving each candidate a chance
depending on his/her share of the vote. It is easy to implement the
lottery in an infallible way, the issue is how its chance should depend
on the share of votes.

In contrast to the present one, the system should avoid any big
jump from a small change in the number of votes. Yet, chance should
not be proportional to the share of votes. Otherwise each voter may
vote for himself, rendering election of a random person. The present
system encourages voters to consolidate around candidates acceptable
to many others. The `jumpless' system should preserve this feature.
This can be done by using a non-linear function: say the chance in the
post-poll lottery be proportional to the {\em squared} number of votes.
In other words, a voter has one vote per each person he agrees
with.\footnote {The dependence of lottery odds on the share of votes
may be sharper.\\ Yet, it must be smooth to minimize the effects of
manipulation. Even (trusty) noise alone,\\ e.g., discarding a randomly
chosen half of the votes, can ``smooth" the system a little.}
 Consider for instance an 8-way race where the percents of votes are 60,
25, 10, 1, 1, 1, 1, 1. The leader's chance will be 5/6, his main rival's
1/7, the third party candidate's 1/43 and the combined chance of the
five `protest' runners 1/866.

This system would force major parties to determine the most popular
candidate via some sort of primaries, and will almost exclude marginal
runners. However it would have no discontinuity rendering any small
change in the vote distribution irrelevant. The system would preserve an
element of chance, but would be resistant to manipulation.

\newpage\section {A Magic Trick}

A book ``Mathematics for Computer Science"\footnote {Problem 15.48
 in a preprint: \hreff {courses.csail.mit.edu/6.042/fall17/mcs.pdf}}
by Eric Lehman, F Thomson Leighton, and Albert R Meyer has
a very nice magic trick with cards. I used in my class some
variation of it described below (with book authors permission).

The trick is performed by a Wizard (W) and his assistant (A) for the viewers
(V).

In W's absence, V choose and give A four cards out of 52 deck.
A places them in a row with one of them ($H$) hidden (turned back up) and exits.
W then enters and guesses $H$.

However, placing $H$ in the middle of the 3 open cards hints that the cards
order is informative, spoiling the surprise. I would instead place the chosen
cards so that, 3 {\bf contiguous} cards are open and 1 hidden, or all are
hidden (sometimes stellar patterns are so favorable to magic that wizards need
no information at all ! \smile.

First, some terms: \trm{Senior} (S), \trm {Junior} (J), \trm {Middle} (M) below
refer to the order of ranks or rank-suit pairs. Kings (K) are special\footnote
 {In Russia, the special one would be Queen, not King:
 Queen of Spades is attributed a special malice. \smile}:
If chosen cards include King of spades~(K0),~all cards are hidden; K1 always
is J, K2 is M, K3 is S. A \trm {4-set} is a set of 4 cards with no K0.

A \trm {string} is an {\em ordered} 4-set with the first or last card replaced
by a symbol $H$ (hidden). $G$ is a bipartite graph of 4-sets connected to four
strings obtained by hiding one card and ordering the rest to reflect the rank
of $H$. A hidden K is treated as a duplicate of the respective (J, M, or S)
non-K open card. The Wizard only needs to figure the suit of $H$.

$G$ breaks into small connected components distinguished by their sets $R$
of non-K ranks of the 4 chosen cards and ranks' multiplicity (including K as
duplicates). With a {\bf uniform degree} 4, $G$ has a {\bf perfect matching},
described below, for A,W to use.

In a 4-set, let $\a$ be the $\Z_4$ sum of all suits in single-suit ranks.
Multiple suits in a rank are viewed in a circle ($\Z_7$ if $|R|{=}1$,
else $\Z_5$) including respective Kings (but not K2 for $|R|{=}2$).
Let $\b$ (and $\b'$ if $2$ such ranks) be $0$ if the suits are consecutive, else
$1$. Notations like $j,j'$ mean same rank suits, $j'{\equiv}j{+}1{+}\b\pmod 5$.
Let $\g$ be $2$ if $|R|{=}2$ with K2 present, else $\g{=}0$.
Below is a simple matching, blind to $\Z_5,\Z_7$ rotations.
(I omit cases with just $j,m,s$ permuted):

\begin{description} \item [$|R|{=}1$] $H$ is the suit in a row (in $\Z_7$)
                   adjacent to $1$-suit-shorter gap (left is preferred).
 \item [$|R|{=}2$] suits $j,j',s,s'$: $H{=}j$ if $\b{=}\b'$, else $H{=}s$.
 \item [$|R|{=}2$] suits $j,c{=}K2,s,s'$ or $j,s,c{=}s',s''$:
             $H{=}j$ if $\a{=}\b{+}\g$; $H{=}c$ if $\a{+}\b{=}1$; else $H{=}s$.
 \item [$|R|{=}3$] suits $j,j',m,s$:
   $H{=}s$ if $x{=}(\a{+}\b \bmod 4$) is $0$; $H{=}m$ if $x{=}1$; else $H{=}j$.
 \item [$|R|{=}4$] The seniority of $H$ reflects $\a$. \end{description}

 \end{document}